\documentclass[12pt]{article}
\usepackage[dvips]{graphicx}
\begin{document}
\title{Decoherence and Programmable Quantum Computation}
\author{Jeff P. Barnes and Warren S. Warren \\
{\it Chemistry Department, Princeton University,
Princeton, NJ 08544-1009} }
\date{ (\today) }
\maketitle
\begin{abstract}
An examination of the concept of using
classical degrees of freedom
to drive the evolution of quantum computers
is given.   Specifically, 
when externally generated, coherent states
of the electromagnetic field are used to
drive transitions within the qubit system, a
decoherence results due
to the back reaction from the qubits onto
the quantum field.
We derive an expression for the 
decoherence rate for two cases, that of the
single-qubit Walsh-Hadamard transform, and 
for an implementation of the controlled-NOT
gate.   We examine the impact of this
decoherence mechanism on Grover's search
algorithm, and on the proposals for use
of error-correcting codes in quantum computation.
\end{abstract}
PACS number(s): 03.67.Lx, 32.80.Qk
%
%
\setcounter{page}{1}
\section{Introduction}
In the original concept of quantum computation,
the isolated, coherent evolution of a
quantum system corresponded to a series of
logical operations which could be used
to compute a solution to a problem \cite{feynman}.
In this concept, once a method of solving a
given problem is decided upon, the logical
steps in the method are translated into unitary
transforms of a quantum system \cite{deutsch}.
These transforms demand a certain form
for the Hamiltonian $H$ that describes the 
evolution of the quantum computer, 
which in turn constrains the architecture of the 
quantum system.  
In other words, the program determines the
system propagator,
$U(t) = \exp(-i H t / \hbar)$,
which determines the form of $H$,
whose parameters indicate the qubit-qubit 
interactions that must be present in order
to carry out the program.
\par
Such quantum computers
suffer from a significant drawback:
they are not programmable.
Consider the case of NMR quantum computing.
A given program to solve a given problem determines
$H$, which in turn determines the magnitudes
of the $J$ couplings between distinct spins,
which in turn constrains the geometry of the
molecule to be used as the quantum system.
Once the molecule is created, 
perhaps utilizing organic synthetic
chemistry, it is in general useful only
for the method of solving the problem originally
decided upon.
In contrast with the flexibility of classical,
transistor-based computers, these
quantum computers have their programs 
``hard-wired'' into their architecture.
\par
However, recent proposals for 
quantum computer architectures
seek to overcome this limitation.
They use external fields, generated by
classical degrees of freedom,
in order to drive the quantum system's evolution
\cite{lloyd}.
Since classical sources can easily be 
manipulated by the (presumably classical) programmer,
these methods offer a means by which the
programmer can alter the evolution of the quantum
system, and thus program the computer.
Some of these proposals include the
use of radio-frequency pulses acting
upon nuclear spins in liquids \cite{jones,cory_II},
laser pulses acting upon ions 
trapped in resonators \cite{cirac,schneider},
electrostatic fields generated 
from gated electrodes influencing
the evolution of nuclear spins 
in semiconductors \cite{kane}, or electrons
trapped in quantum dot structures \cite{chemla}.
\par
There is a question that naturally arises when one
contemplates such proposals:  how can the evolution of
a quantum system remain coherent if it is interacting with
classical degrees of freedom?
Consider the loss of visibility 
of the interference pattern in
a two-slit electron beam experiment 
which occurs when one attempts 
to measure which path the electron took
using a light source \cite{feynman2}.   
The interaction of the field
with the electron creates an entangled state 
of the electron and field, from which
a measurement of the state of the light 
will reveal information about the position
of the electron.  To the extent to which the light
carries information about which path the electron takes,
interference is lost \cite{feynman2,decoherence,zurek}.
Such interference is central to the 
working of many quantum programs, {\it e.g.}
Grover's search algorithm \cite{grover}.
If we use classically generated fields to
drive (and thus interact) with the qubits, 
will that interaction effectively measure the 
state of the qubits, and by doing so, 
destroy the coherent evolution of the system?
\par
To answer this question, we examine
a specific case, that of Grover's quantum
search algorithm \cite{grover}.
First, we review the implementation of
this algorithm when quantum back reaction is
not a concern.   Then we assume that one of
the transforms used in the algorithm,
the Walsh-Hadamard transform, is
driven by externally applied electromagnetic
pulses.   The quantum field is assumed to be
generated by classical sources, so that it
is described by a coherent field state.
Thus, the quantum field exhibits behavior
that is close to a classical field.
Despite this, there is some
quantum back-reaction on the field which leads
to a decoherence of the qubit system.  
We determine
the rate of decoherence due to this process.
Finally, we discuss the broader implications of
this mechanism of decoherence.
While these questions have been previously raised
\cite{imoto}, to the author's knowledge
no quantitative assessment of their importance
has yet been given.
\section{Grover's algorithm}
\subsection{Discussion}
Before we consider how to drive a part of Grover's
search algorithm, let us briefly review it here.
Grover's quantum search algorithm is a method
by which to retrieve elements in a subset 
of a larger set \cite{grover}.
It acts upon $K$ qubits, or two-level systems, 
whose levels are arbitrarily
label as $|0\rangle$ and $|1\rangle$.
A complete, orthonormal basis for the system 
is the product basis, {\it e.g.} for $K$=3, 
$\{ |000\rangle$, $|001\rangle$, $|010\rangle$,
$|011\rangle$, $|100\rangle$, $|101\rangle$,
$|110\rangle$, $|111\rangle \}$.
It is usual to refer to each element of the basis
set by the integer whose binary expansion represents
the string of 0s and 1s, so that 
$|3\rangle \equiv |011\rangle$.
The elements of the set to be searched 
are labeled by the integers from 0 to $2^K-1$.
\par
The subset whose elements we are searching for 
is specified according to a condition.
For example, we could search for any
integer-valued roots of a given polynomials 
within a fixed range.
There are many problems for which, given an $x$ 
of the set of possible solutions, it can be
checked in a polynomial number of steps whether $x$
is a solution to the problem, but no known method
exists to find all solutions in polynomial steps \cite{NP}.
These problems can be solved by a
brute-force search over all possible solutions,
which is what Grover's algorithm does.
Since Grover's algorithm has been shown to be
optimal \cite{bennett,boyer,zalka}, its performance
is one important indicator of how quantum computation
might out-perform classical computation.
\par
In what follows, we assume there is only a single
solution, $y$, to the problem.   It is not difficult
to generalize this to multiple solutions \cite{boyer}.
The initial state of the quantum computer is
$\sum_{x=0}^{2^K-1} |x\rangle / \sqrt{2^K}$.
The goal is to transfer amplitude into
the state $|y\rangle$ \cite{grover} so that
a measurement of the system yields the solution.
This is achieved by a series of transforms of the form
$(\mathcal{W}\mathcal{R}\mathcal{W}\mathcal{O})^N$.
The unitary, Hermitian operator
$\mathcal{W}$ = $\prod_n \mathcal{W}(n)$
= $\prod_n \sqrt{2}$ ($\mathcal{S}_x(n)$
- $\mathcal{S}_z(n)$) is the Welsh-Hadamard transform.
It consists of a product of transforms, acting
independently on each qubit in the system.
(When referring to a single qubit,
we employ the Pauli spin
operator notation as commonly used in
magnetic resonance \cite{ernst}.
In particular, $\mathcal{S}_\alpha$
= $|1\rangle \langle 1|$ and 
$\mathcal{S}_\beta$  = $|0\rangle \langle 0|$
are projection operators, and
$\mathcal{S}_+$ = $|1\rangle \langle 0|$
and $\mathcal{S}_-$ = $|0\rangle \langle 1|$
are raising and lowering operators for
a single qubit.)
\par
The operators $\mathcal{R}$ = $-\mathbf{1}$
+ 2$|0\rangle \langle 0|$
and $\mathcal{O}$ = 
$\mathbf{1}$ - $2|y\rangle \langle y|$
are diagonal in the product basis.
The operator $\mathcal{O}$ is called the oracle.
It is the only means by which the algorithm
has knowledge of the solution.   It would be the
routine the flips the amplitude of the state
$|x\rangle$, for example, if $x$ were the root
of a given polynomial.
The combination $\mathcal{W}\mathcal{R}\mathcal{W}$
can also be written as
$-\mathbf{1}$ + 2$( \sum_{x=0}^{2^K-1} |x\rangle)
(\sum_{x=0}^{2^K-1} \langle x|)/2^K$,
which is called the invert-about-average step.
\par
The algorithm can be understood as 
a combination of two inversions, 
the first about $|y\rangle$ and the second 
about the state with an equal amplitude 
for all basis states.
The two inversions result in a rotation,
transferring amplitude into $|y\rangle$ \cite{grover,zalka}.  
For any normalized state of the computer,
$\sum_{x=0}^{2^K-1} a_x |x\rangle$,
the application of the transformation 
$\mathcal{W}\mathcal{R}\mathcal{W}\mathcal{O}$
alters the state as follows:
\begin{equation}
\left( \begin{array}{c} a_y \\ \\ 
\frac{\displaystyle \sum_{x \ne y} a_x }
{\displaystyle 
\parbox{0.1cm}{\vspace{0.9cm}} \sqrt{2^K-1}}
\end{array} \right)
\stackrel{\mathcal{W}\mathcal{R}\mathcal{W}\mathcal{O}}
{\longrightarrow}
\left( \begin{array}{cc}
1-\frac{\displaystyle 2}
{\displaystyle \parbox{0.1cm}{\vspace{0.7cm}} 2^K} & 
\frac{\displaystyle 2}
{\displaystyle \parbox{0.1cm}{\vspace{0.7cm}} 2^K} 
\sqrt{ \parbox{0.2cm}{\vspace{0.6cm}} \, 2^K-1} \\ \\
-\frac{\displaystyle 2}
{\displaystyle \parbox{0.1cm}{\vspace{0.7cm}} 2^K} 
\sqrt{ \parbox{0.2cm}{\vspace{0.6cm}} \, 2^K-1} & 
1-\frac{\displaystyle 2}
{\displaystyle \parbox{0.1cm}{\vspace{0.7cm}} 2^K}
\end{array} \right)
\left( \begin{array}{c} a_y \\ \\
\frac{\displaystyle \sum_{x \ne y} a_x}
{\displaystyle \parbox{0.1cm}{\vspace{0.9cm}} \sqrt{2^K-1}}
\end{array} \right) 
\label{eqn:rotation} \end{equation}
which is a rotation of the probability amplitude 
between $|y\rangle$ and all other states,
with $\sin \varphi$ $\approx$ $2^{1-K/2}$ for
large $K$.   When $a_y$ and $\sum_{x \ne y} a_x$ have
the same sign, the amplitude for the state $|y\rangle$
increases with every iteration, and decreases otherwise.
\par
Note that $\mathcal{R}$ and $\mathcal{O}$,
unlike $\mathcal{W}$, are not products of
operators acting independently upon each qubit.
They require qubit-qubit interactions to implement.
To see this, write $\mathcal{R}$ =
$-\mathbf{1}$ + 2$\prod_n \mathcal{S}_\beta(n)$
= $-\mathbf{1}$ + 2$\prod_n [ 1/2 - \mathcal{S}_z(n) ]$.
When the product is expanded, terms such as
$\mathcal{S}_z(n) \mathcal{S}_z(m)$ appear,
which indicate the need for qubit interactions.
\subsection{Adding Classical Fields}
Consider how to drive the algorithm utilizing
externally applied fields.   
The Walsh-Hadamard transforms 
$\mathcal{W}(n)$ can be driven one qubit at a time.
Some proposals include methods by which
qubit-qubit couplings,
and thus $\mathcal{R}$ and $\mathcal{O}$,
could also be driven \cite{kane,schneider}.
We assume here that only the $\mathcal{W}(n)$ are
externally driven.  Because of the nature 
of decoherence, it is reasonable
to expect that if further transforms besides $\mathcal{W}(n)$
are driven, the decoherence rate can only 
increase from what is derived below.
\par
Suppose that there exists a 
field / qubit coupling of the 
form $\kappa E(t) \mathcal{S}_x$, where
$E(t)$ is the electric field amplitude,
and $\kappa$ is the field / qubit coupling.
We assume the field uses only a single polarization
for simplicity.   If the qubits are magnetic dipoles, 
the form of the coupling is unchanged, with
$B(t)$ substituted for $E(t)$.
To give our programmer the greatest possible
control, it is usually assumed that 
each qubit can be separately driven by the field.
There are two ways to achieve this: 
spatial resolution (as is usually the case for
lasers acting upon ions in traps) 
or frequency resolution (as is used in magnetic resonance).
In either case, the Fourier components of
the pulses acting upon separate qubits do not overlap.
If spatial resolution is employed, then to prevent
the pulses from overlapping, different directions
of the beams are used.   If frequency resolution
is employed, then the pulses are centered at different
frequencies in order to select a given transition.
\par
The entire qubit / field system is then described
by a Hamiltonian of the form
\begin{equation}
H/\hbar = \sum_n \omega_n \mathcal{S}_z(n)
+ \sum_{n,m} J_{n,m} \mathcal{S}_z(n) \mathcal{S}_z(m)
+ \ldots
+ \kappa \sum_n E_n(\vec{r}_n,t) \mathcal{S}_x(n)
\end{equation} 
where the $J$ terms, and higher-order spin couplings
if necessary, are present to implement
the $\mathcal{R}$ and $\mathcal{O}$ transforms.
The $\mathcal{W}(n)$ transform acting on qubit $n$
at position $\vec{r}_n$ is driven by the field $E_n$.
Assume that the pulses have a square
envelope with a center frequency $\bar{\omega}_n$,
and that $\kappa E_n(\vec{r}_n, t) \gg J_{n,m}$
so that
during a pulse we can ignore the $J$ couplings
(but see note \cite{note1}).
Then implementing $\mathcal{W}$ for each qubit
amounts to finding the propagator for a single
field / qubit interaction acting for a short time,
\begin{equation}
H_n/\hbar = \omega_n \mathcal{S}_z(n) + 
\kappa E_n(\vec{r}_n) 
\cos( \, \bar{\omega}_n t \, ) \mathcal{S}_x(n)
\label{eqn:hi} \end{equation}
We want to adjust these parameters so that we arrive
at the result $\mathcal{W}(n)$ =
$\sqrt{2} (\mathcal{S}_x(n)-\mathcal{S}_z(n))$.
Suppose we write $\mathcal{W}(n)$ =
$\exp(-i \bar{\omega}_n \mathcal{S}_z(n) t \, ) \: U_n(t)$,
which amounts to transforming to a rotating frame
with the field.   Remove the rapidly oscillating terms
to arrive at
\begin{equation}
\frac{\partial}{\partial t}U_n
= -i \left( ( \omega_n - \bar{\omega}_n ) \mathcal{S}_z(n) +
\frac{\kappa}{2} \, E_n(\vec{r}_n) 
\, \mathcal{S}_x(n) \right) U_n
\end{equation}
The propagator for a time-independent
Hamiltonian of the form
$\vec{a} \cdot \vec{\mathcal{S}}$ 
is given by 
$\exp(-i \vec{a}\cdot \vec{\mathcal{S}} t )$
= $\cos (at/2)$ - $2i\sin(at/2)$
$\vec{a}\cdot\vec{\mathcal{S}}/a$.
In our case, 
$\vec{a}$ = $\{ \; \kappa E_n(\vec{r}_n)/2, \;
0, \; \omega_n-\bar{\omega}_n \; \}$,
so to arrive at the correct result we require
the field to be detuned below the qubit,
$\bar{\omega}_n$ = $\omega_n$ - $\Omega/\sqrt{2}$,
with a field strength 
$\kappa E_n(\vec{r}_n) /2$ = $-\Omega/\sqrt{2}$
and a pulse duration of $\pi/\Omega$.
Then
\begin{equation}
\mathcal{W}(n) = i \sqrt{2} \;
\exp \left( -i \pi \frac{\bar{\omega}_n }{\Omega} 
\mathcal{S}_z(n) \right)
\left( \mathcal{S}_x(n) - \mathcal{S}_z(n) \right)
\label{eqn:wh0} \end{equation}
This is the form we seek, except for an extra
phase factor.   Since $\mathcal{W}(n)$ is applied
uniformly to each qubit in the system, if the spread
of frequencies of the separate qubits is not too
large, it will be a constant factor for the entire
quantum computer, and can be ignored.   Otherwise,
suitable delays built into the system can be implemented
to achieve the same purpose.
\section{Inclusion of Quantum Back Reaction}
\subsection{Description of the Initial Field State.}
Before $t=0$, when the computation starts,
the programmer creates pulses of the field
that propagate towards and drive the qubits
at separate times $t_i$.
This is accomplished in a classical manner
by turning on and off a classical current
source, $j(\vec{r},t)$, which interacts with
the field through the vector potential,
$H_P(t)$ = $-\int \, d\vec{r} \, 
\mathcal{A}(\vec{r},t) j(\vec{r},t)$.
For simplicity, a single polarization 
for $j$ is assumed.  The current is classical
in the sense that no quantum back reaction on the
current source was included in the interaction
Hamiltonian.
\par
Grover's algorithm requires a series
of pulses, created at different times, 
call them $t_i$, to drive each qubit $n$.
Let the current distribution that drives qubit $n$
be $j_n (\vec{r},t)$.
It has significant Fourier components over
a range of $\vec{k} \in \mathsf{K}(n)$;
as discussed previously, the $\mathsf{K}(n)$
are mutually disjoint sets.
To create pulse $i$ in the series, 
turn $j_n$ on for a short time $\Delta t$ 
at time $t_i$.
The field state is then transformed by 
$\exp(-i H_P( j_n(\vec{r},t_i) \, ) \Delta t/\hbar)$
= $\prod_{\vec{k} \in \mathsf{K}(n)}$
$D_{\vec{k}} (z_{\vec{k}} \, )$.
The $D_{\vec{k}} (z)$ =
$\exp ( \, z {\mbox {\Large $a$}}_{\vec{k}} {}^\dag
+ z^\star {\mbox {\Large $a$}}_{\vec{k}} \, )$
are called displacement operators. 
They separately transform each mode of the field
according to the complex-valued argument 
$z_{\vec{k}} \propto$ $\Delta t$  
$\int j(\vec{r},t) \exp( \, i \vec{k} \! \cdot \! \vec{r} \, )
\, d\vec{r}$.
The new version of Grover's
algorithm then has the form 
$( \mathcal{W} \mathcal{R} \mathcal{W} \mathcal{O})^N$
$\sum_{x=0}^{2^K-1}|x\rangle$
$\prod_n^{\mathrm{qubits}}$ 
$\prod_{t_i}^{\mathrm{pulses}}$
$\prod_{\vec{k}\in\mathsf{K}(n)}$
$D_{\vec{k}} (z_{\vec{k}}(n,t_i))$
$|\mathrm{vac}\rangle$,
where $|\mathrm{vac}\rangle$ indicates
the vacuum state of the field.
The $\mathcal{W}$ operators are now joint
operators over the qubit and field.
\par
The displacement operators
produce coherent field states 
that have many classical properties.
A list of some of these 
is given in Ref. \cite{mandel}.
We require the following properties:
$D_{\vec{k}} {}^\dag (z) D_{\vec{k}} (z)$ 
= $\mathbf{1}_{\vec{k}}$,
and $D_{\vec{k}} {}^\dag (z)
{\mbox {\Large $a$}}_{\vec{k}} 
D_{\vec{k}} (z)$
= ${\mbox {\Large $a$}}_{\vec{k}}$ + $z$,
and $D_{\vec{k}} {}^\dag (z)
{\mbox {\Large $a$}}_{\vec{k}} {}^\dag
D_{\vec{k}} (z)$ =
${\mbox {\Large $a$}}_{\vec{k}} {}^\dag$
+ $z^\star$.
The above properties lead to the following
useful identity.  Define the positive and 
negative frequency electric field operators
\begin{equation}
\mathcal{E}^+(\vec{r},t) = 
\left( \mathcal{E}^- (\vec{r},t) \right)^\dag =
\sum_{\vec{k}} \: i \,
\sqrt{\frac{\hbar \omega}{2 \epsilon_0 L^3}} \:
{\mbox {\Large $a$}}_{\vec{k}} \, 
\, e^{i \, ( \vec{k} \cdot \vec{r} - \omega t) }
\end{equation}
The total electric field operator,
$\mathcal{E} = \mathcal{E}^+ + \mathcal{E}^-$,
obeys Maxwell's operator equations 
for a source-free region.   
Given a function $f$ that can be 
represented by a Taylor expansion,
then the following holds:
\[ 
f \left( \mathcal{E}^+(\vec{r},t), \:
\mathcal{E}^- (\vec{r},t) \right) 
\exp(-i H_P(j(\vec{r} \, ',t_i) \, ) \Delta t/\hbar)
\]
\begin{equation}
= \exp(-i H_P(j(\vec{r} \, ',t_i) \, ) \Delta t/\hbar) 
f \left( \mathcal{E}^+ (\vec{r},t) + E^+ (\vec{r},t), \:
\mathcal{E}^-(\vec{r},t) + E^-(\vec{r},t) \right)
\label{eqn:commute1} \end{equation}
Although $f$ is still an operator,
it contains the $c$-function $E^+(\vec{r},t)$,
which is the electric field profile that would be 
expected classically at position $\vec{r}$ 
and time $t$, resulting from
a current distribution $j(\vec{r} \, ',t_i)$ 
acting for a time $\Delta t$.
\subsection{The Walsh-Hadamard Transform with a Quantum Field}
The goal of this section is to find a
solution for a single qubit / field interaction,
$\mathcal{W}(n)$.
As before, we assume that the pulses have square
envelopes with center frequency $\bar{\omega}_n$,
and again let the detuning be given by
$\Delta \omega$ = $\omega_n - \bar{\omega}_n$.
We also wish to distribute the
displacement operators into Grover's algorithm
in the following manner:
\[
\prod_n \left\{ \mathcal{W}(n)(t_i) 
\prod_{\vec{k} \in \mathsf{K}(n)} D_{\vec{k}}
(z_{\vec{k}}(n,t_i)) \right\} \,
\mathcal{R} \,
\prod_n \left\{ \mathcal{W}(n)(t_{i-1})
\prod_{\vec{k} \in \mathsf{K}(n)} D_{\vec{k}}
(z_{\vec{k}}(n,t_{i-1})) \right\} \,
\mathcal{O} \cdots
\]
\[
\prod_n \left\{ \mathcal{W}(n)(t_2) 
\prod_{\vec{k} \in \mathsf{K}(n)} D_{\vec{k}}
(z_{\vec{k}}(n,t_2)) \right\} \,
\mathcal{R} \,
\prod_n \left\{ \mathcal{W}(n)(t_1)
\prod_{\vec{k} \in \mathsf{K}(n)} D_{\vec{k}}
(z_{\vec{k}}(n,t_1)) \right\}
\mathcal{O}
\sum_{x=0}^{2^K-1} |x\rangle |\mathrm{vac}\rangle
\]
We have indicated the times at which the
$\mathcal{W}(n)(t_i)$ interaction begins,
{\it e.g.} the time at which the pulse arrives
at the qubit.   In commuting each $D$ past $\mathcal{R}$,
$\mathcal{O}$, and $\mathcal{W}(m)$ with $m \ne n$, 
the $z_{\vec{k}}$ pick up a phase of $\exp(-i \omega t)$.
This corresponds to the free propagation of the
pulse while the other calculations occur.
$D$ acting on qubit $n$ at a later time will
not commute past $\mathcal{W}(n)$ at an earlier
time.   Classically, however, the pulse
has not yet reached the qubit, so the 
expectation value of the $\mathcal{E}_n^\pm$ 
are zero.  The non-zero commuter reflects
the interaction of the qubit with the vacuum, 
including, {\it e.g.} spontaneous emission.
Quantum systems are usually chosen explicitly
to minimize this decoherence mechanism,
so we assume that we can ignore it here.
\par
Start with the interaction Hamiltonian of 
Eq. (\ref{eqn:hi}),
except with the quantum field operators replacing
the classical field.  Transform to frames
rotating with the field and qubit by assuming
a solution of the form
$\mathcal{W}(n)$ = 
$\exp( -i H_0(n) t / \hbar ) W$ where
$H_0(n) / \hbar$ = $\sum_{k \in \mathsf{K}(n)}
\omega \, {\mbox {\Large $a$}}_{\vec{k}} {}^\dag
{\mbox {\Large $a$}}_{\vec{k}}$
+ $\omega_n \mathcal{S}_z(n)$.
This results in an equation for $W$,
\begin{equation}
\frac{d}{dt} W(t) =
-i \frac{\kappa}{2} \left(  
\mathcal{S}_+(n) e^{i \omega_n t} \:
\mathcal{E}_n^+(\vec{r}_n,t)
+ \mathcal{S}_-(n) e^{-i \omega_n t}
\mathcal{E}_n^-(\vec{r}_n,t)
\right) W(t)
\label{eqn:deq1} \end{equation}
The modes over $\mathsf{K}(n)$ in the operator
$\mathcal{E}_n^\pm$ slowly dephase during the
interaction of the pulse with the qubit.
This gives the effect of the pulse envelope on
the qubit, but makes a solution somewhat difficult.
However, our interest is to find only the 
lowest order departures from classical behavior.  
Here is one way to do this:  first, insert a 
unit factor into the propagator,
$\exp(-i H_0(n) t/\hbar)$ 
$\prod_{\mathsf{K}(n)}
D_{\vec{k}} D_{\vec{k}} {}^\dag$
$W(t)$ $\prod_{\mathsf{K}(n)} 
D_{\vec{k}}$ $|\mathrm{vac}\rangle$.
The $D$ to the right commutes to the front
of the entire algorithm, and we solve
for $U \equiv D^\dag W D$ instead.
From Eq. (\ref{eqn:commute1}),
this results in replacing $\mathcal{E}_n^\pm$
with $\mathcal{E}_n^\pm + E_n^\pm$ 
in Eq. (\ref{eqn:deq1}), where $E_n^\pm$
is the classical field at the qubit.
\par
Two things have changed.
First, the appearance of the classical
field profiles in Eq. (\ref{eqn:deq1})
means that part of $U$ alters the qubit evolution
according to the classical field profile.
Photons are still absorbed and emitted by the
qubit, but the process occurs in such a way so that
the coherent state of the field is not altered
(which is a very ``classical'' behavior).
Second, $U$ still contains field operators, 
but now they act upon the vacuum state of the field
with the $D$ to the left.
$U$ will create one-photon states (and higher
orders as well) which are
orthogonal to the vacuum state.
Because $D$ is a unitary transform, 
the displaced one-photon
states are orthogonal to the displaced vacuum
states that describe the pulses.
Thus, $U$ can entangle the qubit state 
with a new field state orthogonal to the original
field state.   This is precisely 
the description of a decoherence mechanism.
\par
The idea that the classical field represents
the lowest-order behavior suggests we should
expand out $U$ as a series, $\sum_j U_j$,
for which $j$ corresponds to the power
of the quantum field operators 
(or equivalently, the power of $\sqrt{\hbar}$).
This results in
\[
\frac{d}{dt}U_0 = -i \frac{\kappa}{2}
\left( E_n^+ \mathcal{S}_+(n)e^{i \omega_n t}
+ E_n^- \mathcal{S}_-(n) e^{-i \omega_n t}
\right) U_0
\]
for the classical behavior, and for $j \ge 1$,
\[
\frac{d}{dt} U_j = -i \frac{\kappa}{2}
\left( E_n^+ \mathcal{S}_+(n) e^{i\omega_n t}
+ E_n^- \mathcal{S}_-(n) e^{-i\omega_n t} \right) U_j
-i \frac{\kappa}{2} \left(
\mathcal{E}_n^+ \mathcal{S}_+(n) e^{i\omega_n t}
+ \mathcal{E}_n^- \mathcal{S}_-(n) e^{-i\omega_n t}
\right) U_{j-1}
\]
$U_1$ incorporates the field operators
once, so it represents the lowest order 
quantum effects.
Each $U_j$ has a solution of the form
$\mathcal{S}_\alpha U_{\alpha,j}(t)$ + 
$\mathcal{S}_\beta U_{\beta,j}(t)$ +
$\mathcal{S}_+ U_{+,j}(t)$ + 
$\mathcal{S}_- U_{-,j}(t)$,
\[
\frac{d}{dt} U_{\alpha,0} = 
-i \frac{\kappa}{2} \, e^{i \omega_n t} 
E_n^+ U_{-,0}
\; \; \; \;
\frac{d}{dt}U_{\beta,0} = -i \frac{\kappa}{2} \, 
e^{-i \omega_n t} E_n^- U_{+,0}
\]
\begin{equation}
\frac{d}{dt} U_{+,0} = -i \frac{\kappa}{2} \,
e^{i \omega_n t} E_n^+ U_{\beta,0}
\; \; \; \;
\frac{d}{dt} U_{-,0} = -i \frac{\kappa}{2} \,
e^{-i \omega_n t} E_n^- U_{\alpha,0}
\label{eqn:deq2} \end{equation}
and for $j \ge 1$,
\[
\frac{d}{dt} U_{\alpha,j} = 
-i \frac{\kappa}{2} \, e^{i \omega_n t} 
\left( E_n^+ U_{-,j} +
\mathcal{E}_n^+ U_{-,j-1} \right)
\; \; \; \;
\frac{d}{dt}U_{\beta,j} = -i \frac{\kappa}{2} \,
e^{-i \omega_n t} 
\left( E_n^- U_{+,j} + \mathcal{E}_n^- U_{+,j-1}
\right)
\]
\begin{equation}
\frac{d}{dt} U_{+,j} = -i \frac{\kappa}{2} \, 
e^{i \omega_n t} \left(
E_n^+ U_{\beta,j} + 
\mathcal{E}_n^+ U_{\beta,j-1} \right)
\; \; \; \;
\frac{d}{dt} U_{-,j} = -i \frac{\kappa}{2} 
e^{-i \omega_n t} \left(
E_n^- U_{\alpha,j} +
\mathcal{E}_n^- U_{\alpha,j-1} \right)
\label{eqn:deq3} \end{equation}
At this point we add in the classical field,
whose envelope is a square pulse.
Thus, let $E_n^\pm$ = 
$E e^{\mp i \phi} e^{\mp i \bar{\omega} t}$, 
where $\phi$ is the phase, and $2E$ 
is the field intensity.
Let $\mathcal{L}_\pm$ =
$(d/dt)^2 \pm i \Delta \omega (d/dt) + (\kappa E/2)^2$
be linear differential operators.
We can then rearrange Eq. (\ref{eqn:deq2}) and
(\ref{eqn:deq3}) to read
\begin{equation}
\mathcal{L}_- U_{\alpha,0} = 0, \; \; \; \;
\mathcal{L}_+ U_{\beta,0} = 0, \; \; \; \;
\mathcal{L}_- U_{+,0} = 0, \; \; \; \;
\mathcal{L}_+ U_{-,0} = 0
\label{eqn:deq4} \end{equation}
and for $j \ge 1$,
\[
\mathcal{L}_- U_{\alpha,j} =
-i\frac{\kappa}{2}
\left( \frac{d}{dt} - i \Delta \omega \right)
\left( e^{i \omega_n t}
\mathcal{E}_n^+ U_{-,j-1} \right)
- \left( \frac{\kappa}{2} \right)^2
E e^{-i\phi} e^{-i \bar{\omega}_n t}
\mathcal{E}_n^- U_{\alpha,j-1}
\]
\[
\mathcal{L}_+ U_{\beta,j} =
-i\frac{\kappa}{2}
\left( \frac{d}{dt} + i\Delta \omega \right)
\left( e^{-i\omega_n t}
\mathcal{E}_n^- U_{+,j-1} \right)
- \left( \frac{\kappa}{2} \right)^2
E e^{i\phi} e^{i \bar{\omega}_n t}
\mathcal{E}^+ U_{\beta,j-1}
\]
\[
\mathcal{L}_- U_{+,j} =
-i\frac{\kappa}{2}
\left( \frac{d}{dt} - i\Delta \omega \right)
\left( e^{i\omega_n t}
\mathcal{E}_n^+ U_{\beta,j-1} \right)
- \left( \frac{\kappa}{2} \right)^2
E e^{-i\phi} e^{-i \bar{\omega}_n t}
\mathcal{E}_n^- U_{+,j-1}
\]
\begin{equation}
\mathcal{L}_+ U_{-,j} =
-i\frac{\kappa}{2}
\left( \frac{d}{dt} + i\Delta \omega \right)
\left( e^{-i\omega_n t}
\mathcal{E}_n^- U_{\alpha,j-1} \right)
- \left( \frac{\kappa}{2} \right)^2
E e^{i\phi} e^{i \bar{\omega}_n t}
\mathcal{E}_n^+ U_{-,j-1}
\label{eqn:deq5} \end{equation}
The initial conditions are
$U_{\alpha,0}$ = $U_{\beta,0}$ = 1,
and $U_{\pm,0}$ = 0, and $U_j$ = 0 for
all $j \ge 1$; and for the first
derivatives,
$d U_{\alpha,0} /dt$ = 
$d U_{\beta,0} /dt$ = 0,
and $d U_{\pm,0} /dt$ = 
$-i (\kappa/2) E e^{\mp i \phi}$,
and $d U_{\alpha,1}/dt$ =
$d U_{\beta,1}/dt$ = 0,
and $d U_{\pm,1}/dt$ =
$-i(\kappa/2)\mathcal{E}_n^\pm(\vec{r}_n,0)$,
and $d U_j / dt$ = 0 for all
$j \ge 2$.
The solutions for $j$ = 0 are
\[
U_{\alpha,0} = e^{i \Delta \omega t /2}
\left[ \cos \theta t
- i \: \frac{\sin \theta t}{\theta}
\frac{\Delta \omega}{2} \right]
\; \; \; \;
U_{\beta,0} = e^{-i \Delta \omega t /2}
\left[ \cos \theta t
+ i \: \frac{\sin \theta t}{\theta}
\frac{\Delta \omega}{2} \right]
\]
\begin{equation}
U_{+,0} = -i e^{i \Delta \omega t / 2}
\left[ \: \frac{\sin \theta t}{\theta} \right]
\frac{\kappa}{2} E e^{-i\phi}
\; \; \; \;
U_{-,0} = -i e^{-i \Delta \omega t / 2}
\left[ \: \frac{\sin \theta t}{\theta}
\right] \frac{\kappa}{2} E e^{i\phi}
\label{eqn:sol1} \end{equation}
where the tip angle is given by
$\theta$ = $\sqrt{ \Delta \omega^2+ (\kappa E)^2 } \, /2$.
This is a restatement of the usual expression
for a Bloch vector influenced by a monochromatic field.
\par
The solutions for $j$ = 1 are linear
in the creation and annihilation operators.
However, the field state they operate on
is the vacuum state, so for the lowest order
effect we will only require the
solution for the creation operators.
They are reasonably complex, so we simplify by
setting $\Delta \omega$ = $\Omega/\sqrt{2}$, 
$\kappa E$ = $-\Omega/\sqrt{2}$,
and the pulse length to $\pi/\Omega$
to reproduce the classical Walsh-Hadamard
transform.   Then, to lowest order,
and still in the doubly rotating frame,
our transform is given by
\begin{equation}
\frac{i}{\sqrt{2}}
\left( \begin{array}{cc}
e^{-i\pi/\sqrt{8}} & 
e^{-i\pi/\sqrt{8}}e^{i\phi} \\ 
e^{i\pi/\sqrt{8}}e^{-i\phi} & 
-e^{i\pi/\sqrt{8}} \end{array} \right)
+
\frac{\kappa}{4\Omega} 
\sum_{\vec{k} \in \mathsf{K}(n)}
\sqrt{\frac{\hbar \omega}{2\epsilon_0 L^3}}
\left( \begin{array}{cc}
e^{-i\pi/\sqrt{8}} e^{-i\phi} g_\beta &
e^{-i\pi/\sqrt{8}} g_- \\
e^{i\pi/\sqrt{8}} e^{2i\phi} g_+ &
e^{i\pi/\sqrt{8}} e^{-i\phi} g_\alpha 
\end{array} \right)
e^{-i \vec{k} \cdot \vec{r}_n }
{\mbox {\Large $a$}}_{\vec{k}} {}^\dag
\label{eqn:wh1} \end{equation}
The functions that give the Fourier
components for the shape of the one-photon
``back-reaction field'' are given by
\[
g_\alpha = \frac{\displaystyle 
2x^2 + \frac{x}{\sqrt{2}} - 1 +
\left(
1 + \frac{x}{\sqrt{2}}
\right) e^{i\pi x} }
{\displaystyle 2\sqrt{2} \: x \: (x^2 -1) }
\; \; \; \; \; \;
g_\beta = -\frac{\displaystyle
1 + \frac{x}{\sqrt{2}} +
\left(
2x^2 + \frac{x}{\sqrt{2}} - 1
\right) e^{i\pi x} }
{\displaystyle 2\sqrt{2} \: x \: (x^2-1) }
\]
\[
g_+ = -\frac{\displaystyle 1 + e^{i\pi x} }
{\displaystyle 4(x^2-1)}
\; \; \; \; \; \;
g_- = \frac{\displaystyle
\left( 4x + 3\sqrt{2} \: \right)
\left( 1+e^{i\pi x} \right) }
{\displaystyle 4\sqrt{2}(x^2-1)}
\]
for which 
$x \equiv (\omega - \bar{\omega}_n)/\Omega$,
the offset of a field mode from the center frequency
of the pulse, in units of the detuning.
\par
\begin{figure}[htpb] \centering
\includegraphics[angle=0,width=5in,height=3.3in]
{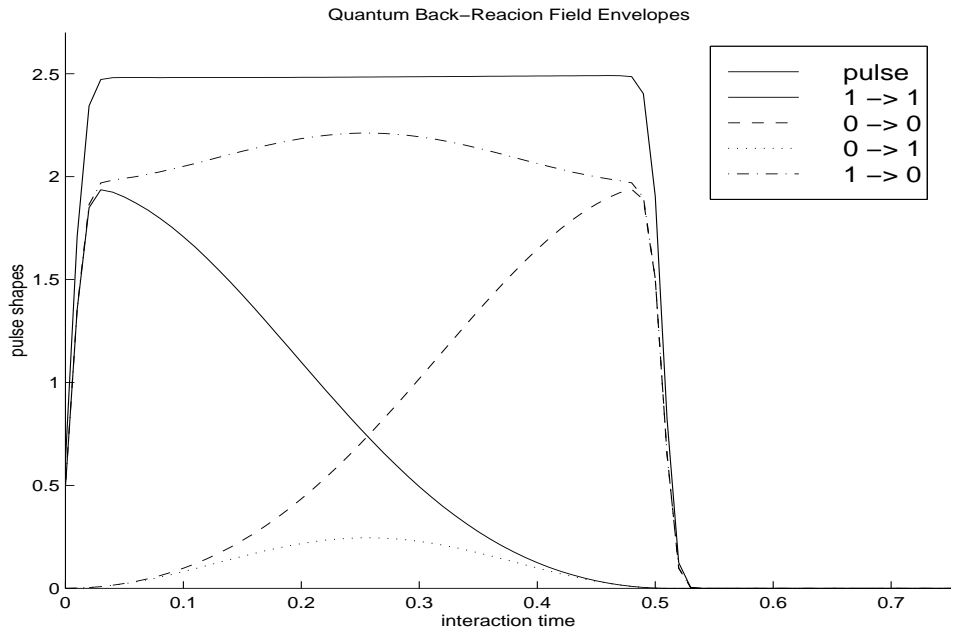}
\caption{FIG. 1. A plot of the envelope of the field that
describes the one-photon state emitted by a qubit
whose state changes as indicated.   For comparison,
the envelope of the classical pulse is also shown.
Time is in units of $\Omega$.}
\label{fig:graph} \end{figure}

The system is assumed to be contained in a
resonator of size $L^3$.   The resonator is
not perfect, but has a finite bandwidth associated
with it due to coupling and conductive losses,
which allows for the passage of the classical
square pulse (which has the Fourier components
$(e^{i\pi x}-1)/x$ ).
Because all the $g$ functions have significant
Fourier components near $\bar{\omega}$
and decay as $x^{-1}$ like the square pulse, 
the resonator also allows the one-photon states 
created by the qubit to escape.   
The time-domain envelopes of these 
one-photon states
(the Fourier transform of the $g$ functions)
are shown in Fig. \ref{fig:graph}.
\par
As in the classical field case, 
the extra phase factors of $\pm i\pi/\sqrt{8}$
in the transformation Eq. (\ref{eqn:wh1}) can be
dropped, since they can be 
compensated for by proper timing.
Similarly, we assume $\phi$ = 0 at the
start of the interaction.   The $\phi$ term
does indicate a curious feature of the
back reaction: $g_-$ contains 
no term from the classical field, 
while $g_+$ requires two interactions with
the classical field in order to perturb 
the coherent state.   This is consistent with
the property that the removal of a photon
from a coherent state does not alter it, while
the addition of a photon does.
\par
We can rewrite the transform Eq. (\ref{eqn:wh1})
in terms of unitary field operators,
\begin{equation}
\frac{i}{\sqrt{2}} 
\left( 
\begin{array}{cc} 1 & 1 \\ 1 & -1 \end{array} 
\right) + 
\frac{\kappa}{4\Omega} 
\sqrt{\frac{\hbar \bar{\omega}_n}{2 \, \epsilon_0 L^3} }
\left( \begin{array}{cc} 
\sqrt{I_\beta} \: G_\beta &
\sqrt{I_-} \: G_- \\ 
\sqrt{I_+} \: G_+ & 
\sqrt{I_\alpha} \: G_\alpha
\end{array} \right)
\label{eqn:wh2} \end{equation}
where the $G$ operators create normalized
one-photon field states, so that
$G^\dag G$ = $\mathbf{1}$.
The normalization factors were found by
numerical integration, for which the slowly
varying $\sqrt{\omega}$
term was ignored: 
$I_\alpha$ =
$\int_{-\infty}^{\infty} |g_\alpha(x)|^2 dx$
= 4.297, 
$I_\beta$ = 4.297,
$I_+$ = 0.617, 
and $I_-$ = 10.451.
Although the $G|\mathrm{vac}\rangle$
states are orthogonal to the initial 
coherent state, they are not mutually orthogonal.
Later on, we will require their (non-normalized) 
overlap integrals
$I_{\alpha,\beta}$ =
$\sqrt{I_\alpha I_\beta}
\langle \mathrm{vac}|G_\alpha^\dag G_\beta
|\mathrm{vac}\rangle$ =
$\int_{-\infty}^{\infty} g_\alpha^\star(x)
g_\beta(x) dx$
= $0.614+i2.221$,
$I_{\alpha,+}$ = $-0.617+i1.110$,
$I_{\alpha,-}$ = $4.300-i3.331$,
$I_{\beta,+}$ = $0.617+i1.110$,
$I_{\beta,-}$ = $-4.300-i3.331$,
and $I_{+,-}$ = $-1.850$.
\par
The important result of this section
is the transformation, Eq. (\ref{eqn:wh2}).   
Each qubit / field interaction 
has a probability amplitude,
proportional to $\Omega^{-1}$, to entangle
the qubit with a field state orthogonal to the
original field state.   The greater $\Omega$,
the lesser the decoherence via this mechanism.
Actually, this may seem counterintuitive.
Consider the interference pattern 
produced by a coherent beam of electrons 
incident upon a double slit.
Now allow a laser to interact with one of the two
paths the electron could travel from the slit
to the detector.  If photons are scattered out of
the coherent modes into vacuum states, then
the visibility of the interference pattern is degraded,
as expected \cite{feynman2}.
If, however, only stimulated processes are important, 
then the visibility of the interference pattern 
should increase as the laser intensity is increased.
The Poisson statistics of a
coherent state can more efficiently
hide the information about which
path the electron takes as the
number of photons in the beam increases.
A similar situation has been noted with regard to
{\it welcher Weg} experiments in atomic 
interferometry (see {\it e.g.} \cite{scully}).
\subsection{Grover's Algorithm with a
Quantum Walsh-Hadamard Transform}
Taking the result from the previous section,
Grover's algorithm is now
\begin{equation}
\prod_{j}^{\mathrm{steps}} \bigg[
\prod_{n=1}^K \left( \mathbf{1} -
\frac{i\kappa}{8\Omega}
\sqrt{\frac{\hbar \bar{\omega}_n}{\epsilon_0 L^3}}
\mathcal{A}(n,t_{j+1})
\right)
\mathcal{W}\mathcal{R}\mathcal{W} 
\prod_{n=1}^K \left( 
\mathbf{1} -
\frac{i\kappa}{8\Omega}
\sqrt{ \frac{\hbar \bar{\omega}_n }{\epsilon_0 L^3} }
\mathcal{B}(n,t_j) \right)
\mathcal{O} \; \bigg] \;
\left( \frac{1}{\sqrt{2^K}}
\sum_x |x\rangle \right) 
|\mathrm{vac}\rangle
\label{eqn:newgrover} \end{equation}
The displacement operators are not shown.
They have been commuted all the way left,
picking up the appropriate phase factors 
representing the propagation of the free field.
The $\mathcal{A}$ and 
$\mathcal{B}$ are from transform Eq. (\ref{eqn:wh2}),
from which a factor of $\mathcal{W}(n)$ 
is first removed:
\begin{equation}
\mathcal{A} =
\left( \begin{array}{cc} 
\sqrt{I_\beta}G_\beta + \sqrt{I_-}G_- &
\sqrt{I_\beta}G_\beta - \sqrt{I_-}G_- \\
\sqrt{I_+}G_+ + \sqrt{I_\alpha}G_\alpha &
\sqrt{I_+}G_+ - \sqrt{I_\alpha}G_\alpha
\end{array} \right)
\; \; \; \;
\mathcal{B} =
\left( \begin{array}{cc}
\sqrt{I_\beta} G_
\beta + \sqrt{I_+} G_+ &
\sqrt{I_-}G_- + \sqrt{I_\alpha}G_\alpha \\
\sqrt{I_\beta} G_\beta - \sqrt{I_+}G_+ &
\sqrt{I_-} G_- - \sqrt{I_\alpha} G_\alpha
\end{array} \right) 
\label{eqn:ab} \end{equation}
The fact that operators acting independently
upon separate qubits and field modes mutually
commute was also used.
\par
When Eq. (\ref{eqn:newgrover}) is expanded out
and only the lowest order terms are kept,
to the original Grover's transform are added
new terms.   
As previously discussed, the field operators
in front of each term, one for each qubit and
each pulse, create
field states that are 
both mutually orthogonal,
and orthogonal to the original field state.
Thus, the probability that
the final qubit state is $|y\rangle$
is the sum of the probabilities
for each of separate terms to be in $|y\rangle$.
In order to preserve normalization to 
$\mathcal{O}(\Omega^{-2})$,
the total probability to be
in an orthogonal field state 
is subtracted from the probability of
finding the field still in the original
state at the end of the algorithm.
The goal of this section is to calculate the
success of this implementation of Grover's algorithm.
\par
After $j$ successful steps of the algorithm,
the computer state is given by
\[
\left( \: \frac{\displaystyle
\cos ( j\varphi ) }{\displaystyle
\sqrt{2^K-1} } \:
\sum_{x \ne y}
|x\rangle + 
\sin ( j\varphi ) \: |y\rangle \right) \:
|\mathrm{vac}\rangle
\]
where $\sin \varphi$ = 
$2 \sqrt{2^K-1}/2^K$ \cite{zalka}.
The general trend for the influence
of the back reaction can be discerned
from the specific example of 3 qubits
and $y$ = 2.
A back reaction on the field can occur
during either application of $\mathcal{W}$.
If it occurs immediately after 
$\mathcal{O}$ has been called (the operator
$\mathcal{B}$), 
or after the invert about average transform
(the operator $\mathcal{A}$).
Suppose the reaction occurs 
after $\mathcal{O}$, and it is for
the least significant qubit.   
Then the computer state becomes 
\[
\Bigg\{
\frac{\cos (j\varphi) }{\sqrt{2^K-1}}
\Big( \:
(B_\beta |0\rangle + B_+ |1\rangle) +
(B_\alpha |1\rangle + B_- |0\rangle ) +
(B_\alpha |3\rangle + B_- |2\rangle ) +
(B_\beta |4\rangle + B_+ |5\rangle ) +
\]
\begin{equation}
(B_\alpha |5\rangle + B_- |4\rangle ) +
(B_\beta |6\rangle + B_+ |7\rangle ) +
(B_\alpha |7\rangle + B_- |6\rangle )
\: \Big) -
\sin (j\varphi) \: \Big( \:
B_\beta |2\rangle + B_+ |3\rangle \: \Big)
\Bigg\} |\mathrm{vac}\rangle
\label{eqn:first_qubit} 
\end{equation}
The amplitudes of pairs of states that differ at
their least significant digit such
as (0,1), (2,3), and so on, are mixed.
If the error occurs for the second least 
significant qubit, then
\[ \Bigg\{
\frac{\cos (j\varphi) }{\sqrt{2^K-1}}
\Big( \:
(B_\beta |0\rangle + B_+ |2\rangle) +
(B_\alpha |1\rangle + B_- |3\rangle ) +
(B_\alpha |3\rangle + B_- |1\rangle ) +
(B_\beta |4\rangle + B_+ |6\rangle ) +
\]
\begin{equation}
(B_\alpha |5\rangle + B_- |7\rangle ) +
(B_\beta |6\rangle + B_+ |4\rangle ) +
(B_\alpha |7\rangle + B_- |5\rangle )
\: \Big) -
\sin (j\varphi) \: \Big(
B_\beta |2\rangle + B_+ |0\rangle \Big)
\Bigg\} |\mathrm{vac}\rangle
\label{eqn:second_qubit}
\end{equation}
The difference is in which pairs of states
are mixed, and whether the qubit involved
in the back reaction was initially in
state 0 or 1.   Using this, we can sum 
over all the qubits of the computer, the
total probability for a back reaction to occur
during the first of the two applications
of $\mathcal{W}$, at step $j$ in the algorithm.
Suppose that the qubit
frequencies are sufficiently close so that all
pulse center frequencies are $\bar{\omega}$.
Then
\[
\left( \frac{\kappa}{8\Omega} \right)^2
\frac{\hbar \bar{\omega} }{\epsilon_0}
\Bigg[
\frac{K}{2} \frac{2^K-2}{2^K-1} \cos^2 (j\varphi)
\bigg( | (B_\beta + B_-)|\mathrm{vac}\rangle |^2
+ | (B_\alpha + B_+)|\mathrm{vac}\rangle |^2 \bigg) + 
\]
\[
(K - \| y \|) \Big(
\left| \frac{\cos (j\varphi)}{\sqrt{2^K-1}} 
B_- |\mathrm{vac}\rangle - 
\sin (j\varphi) \: B_\beta |\mathrm{vac}\rangle 
\right|^2 +
\left| \frac{\cos (j\varphi)}{\sqrt{2^K-1}} 
B_\alpha |\mathrm{vac}\rangle - 
\sin (j\varphi) \: B_+|\mathrm{vac}\rangle \right|^2
\Big) + 
\]
\[
\| y \| \Big( \:
\left| \frac{\cos (j\varphi)}{\sqrt{2^K-1}} 
B_\beta |\mathrm{vac}\rangle -
\sin (j\varphi) \: B_- |\mathrm{vac}\rangle \right|^2 +
\left| \frac{\cos (j\psi)}{\sqrt{2^K-1}} 
B_+ |\mathrm{vac}\rangle -
\sin (j\varphi) \: 
B_\alpha |\mathrm{vac}\rangle \right|^2 \Big)
\Bigg]
\]
The factor $\| y \|$ appears since the
back reaction depends upon whether a qubit
was initially in state 0 or 1.
Now suppose $K$ is large enough so that
$2^K \gg 1$.   
The following approximations are useful:
$\sum_{j=1}^{\pi/2\varphi} \cos^2(j\varphi)$
$\approx$ $\int_{0}^{\pi/2} \cos^2 x \: dx / \varphi$
= $\pi/4\varphi$, and
$\sum_{j=1}^{\pi/2\varphi} \sin^2(j\varphi)$
$\approx$ $\pi/4\varphi$, and
$\sum_{j=1}^{\pi/2\varphi} \sin(j\varphi) \cos(j\varphi)$
$\approx$ $1/2\varphi$.
Again, keeping only the largest
terms in $K$,
the probability that a back reaction occurs
for any qubit, and at any step,
during the first of the two
$\mathcal{W}$ transforms
is given by
\[
\left( \frac{\kappa}{8\Omega} \right)^2
\frac{\hbar \bar{\omega}}{\epsilon_0}
\frac{\pi}{8} 2^{K/2}
\Bigg[ \frac{K}{2} \Big(
\big| (B_\beta + B_+)|\mathrm{vac}\rangle \big|^2 +
\big| (B_\alpha + B_+)|\mathrm{vac}\rangle \big|^2
\Big)
\]
\begin{equation}
+ ( K - \| y \| )
\Big( 
\big| B_\beta |\mathrm{vac}\rangle \big|^2 +
\big| B_+ |\mathrm{vac}\rangle \big|^2
\Big)
+ \| y \| 
\Big( 
\big| B_- |\mathrm{vac}\rangle \big|^2 +
\big| B_\alpha |\mathrm{vac}\rangle \big|^2
\Big)
\Bigg]
\label{eqn:firstdecohere} \end{equation}
If the back reaction occurs during the second
$\mathcal{W}$ in an iteration, then $j$ starts at
2 (the first invert-about-average step is always
carried out), and the sign of the amplitude
for $|y\rangle$ is positive.   Taking the limit
for large $K$, the cross-terms that depend upon
the sign of $a_y$ drop out, and the final expression
is the same as above except with the $B$ operators
replaced by those of $A$.
The total probability of a back reaction
anywhere in the algorithm is then the sum of
these two.   Starting with the operators of
Eq. (\ref{eqn:ab}),
the matrix elements are expressed in terms of
the normalization and overlap of
the $G|\mathrm{vac}\rangle$ states,
{\it e.g.} 
$|(B_\beta +B_- )|\mathrm{vac}\rangle|^2$
+ $|(B_\alpha +B_+ )|\mathrm{vac}\rangle |^2$
= $2(I_\alpha + I_\beta + I_+ + I_-)$
+ $4 \Re e ( I_{\beta,-} + I_{\alpha,+} )$.
Plugging in, the final probability
to end up entangled with an orthogonal field
state is
\begin{equation}
\left( \frac{\kappa}{\Omega} \right)^2
\frac{\hbar \bar{\omega}}{\epsilon_0}
\sqrt{2^K}
\bigg( 0.211 K + 0.241 \| y \| \bigg).
\label{eqn:decohere} \end{equation}
\par
How much do the orthogonal field states
contribute to the correct final answer?
First, note that the matrix elements of 
$\mathcal{A}$ and $\mathcal{B}$
all have similar magnitudes.
Thus, they equally mix the state $|y\rangle$
with the state connected to it by flipping the
qubit that experiences the back reaction.
Early in the algorithm when $\sin( j \varphi) \ll 1$,
this does not increase the amplitude 
in $a_y$ significantly, since the state that 
mixes with $|y\rangle$ has amplitude 
$\cos( j \varphi)/\sqrt{2^K} \ll 1$.
At later times, however, the amplitude of 
$|y\rangle$ is near 1, so the back reaction
decreases the probability to be 
in state $|y\rangle$ by roughly half.
\par
Thus, after $j$ iterations of Grover's algorithm,
a back reaction will cause the result to appear
as if only $\approx j/2$ iterations had been
performed.   Further iterations will continue
to transform these entangled states, 
but this does not
necessarily improve the final result.
Recall from Eq. (\ref{eqn:rotation})
that amplitude is rotated into
$|y\rangle$ only when the signs 
of $a_y$ and $\sum_{x \ne y}|x\rangle$
are the same.   For large $K$, the
amplitude $a_y$ is $\approx \sin(j\varphi)$
$( \sqrt{I_\beta}G_\beta
+ \sqrt{I_-}G_-) |\mathrm{vac}\rangle$.
The other qubit states are entangled with
field states that are partly orthogonal 
to this state, but the amplitude that lies
along the same direction in the Hilbert space
of the field is, for large $K$,
$\cos (j\varphi)$ $( I_\beta$ +
$I_{\beta,+}$ + $I_{-,\beta}$ + $I_{-,+})$
$/\sqrt{I_\beta + I_- + 2 \Re e(I_{\beta,-})}$
= $(-2.232+i0.448)\cos (j\varphi)$.
The real part has switched sign, and so further
iterations will actually remove amplitude
from $|y\rangle$.   It should be clear that
the back reaction essentially scrambles the memory
of the quantum computer, from which further
iterations will not, in general, recover the
correct amplitude before the computation ends.
\section{Quantum Back Reaction for the CNOT Gate}
In this section, we briefly examine what
happens when a classical field is used
to drive a logic gate, such as $\mathcal{R}$,
that entangles together separate qubits.
As before, let us focus on a
single example, that of the controlled-NOT
(CNOT) gate.   CNOT is defined as a transform 
acting on two qubits that flips the second qubit
only if the first qubit is 1:
$|00\rangle \rightarrow |00\rangle$,
$|01\rangle \rightarrow |01\rangle$,
$|10\rangle \rightarrow |11\rangle$, and
$|11\rangle \rightarrow |10\rangle$.
As for the Walsh-Hadamard transform,
there are numerous ways to implement the
CNOT gate \cite{cory_II}.   Perhaps the
most straight forward method is the use
of a selective pulse.
What we mean by this is the following.
\par
The two qubits to be transformed by
CNOT have a Hamiltonian
\begin{equation}
H/\hbar = 
\omega_1 \mathcal{S}_z + \omega_2 \mathcal{I}_z
+ J \vec{\mathcal{S}} \cdot \vec{\mathcal{I}} +
\sum_{\vec{k}} \omega 
{\mbox {\Large $a$}}_{\vec{k}} {}^\dag
{\mbox {\Large $a$}}_{\vec{k}}
+ \kappa 
\Big( \mathcal{S}_x + \mathcal{I}_x \Big)
\Big( \mathcal{E}^+ + \mathcal{E}^- \Big).
\label{eqn:ham1} \end{equation}
By $\vec{\mathcal{S}}\cdot\vec{\mathcal{I}}$
we mean the qubit operator
$\mathcal{S}_x \mathcal{I}_x$+
$\mathcal{S}_y \mathcal{I}_y$+
$\mathcal{S}_z \mathcal{I}_z$.
Assume that
$\omega_2 > \omega_1$ $\gg$ 
$|\omega_1 - \omega_2|$ $\gg$ 
$J$.   In this case, the
transitions driven by the field
occur at $\approx$
$\omega_1 \pm J/2 + J^2/4(\omega_1-\omega_2)$
and $\approx$ 
$\omega_2 \pm J/2-J^2/4(\omega_1-\omega_2)$,
They appear as a ``doublet of doublets'', 
that is, two sets of two lines each.
Since $H$ is nearly diagonal for small $J$,
the highest frequency transition corresponds to 
$|10\rangle \leftrightarrow |11\rangle$.
Thus, a CNOT gate can be implemented by
selectively inverting this transition
with a square pulse of duration $T \gg 2\pi/J$.
($T$ also depends on the matrix element
for the transition.)
The long length of the pulse keeps its
bandwidth small enough so that it does
not significantly overlap with the other
transitions in the system.
\par
\begin{figure}[htpb] \centering
\includegraphics[angle=0,width=5in,height=3.3in]
{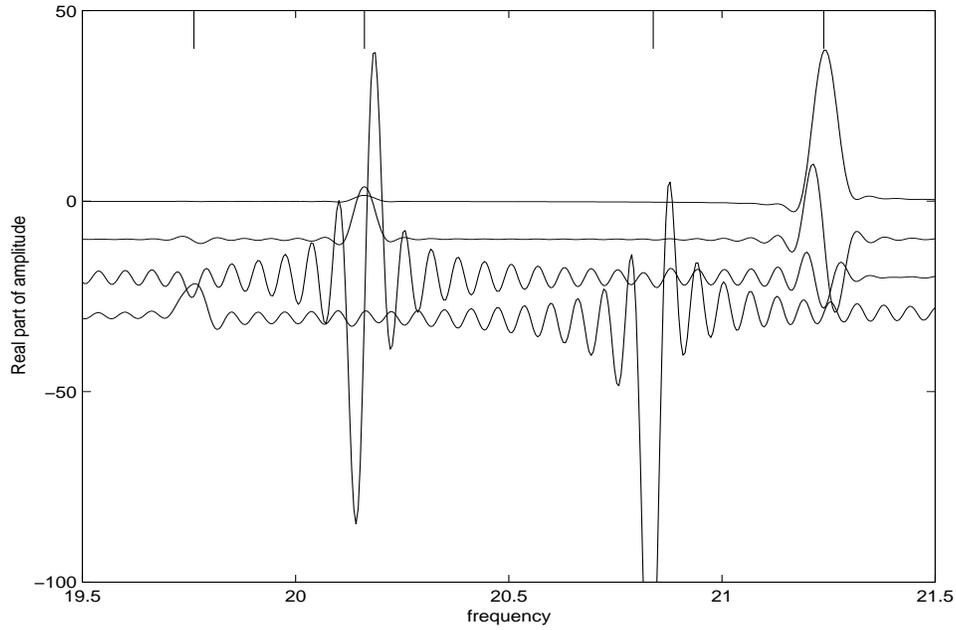}
\caption{FIG 2.  
A plot of the real part of the
frequency-dependent matrix elements
for transitions (from top to bottom)
$|11\rangle \rightarrow |11\rangle$,
$|11\rangle \rightarrow |10\rangle$,
$|10\rangle \rightarrow |01\rangle$, and
$|01\rangle \rightarrow |00\rangle$.
The lines at top represent the transitions
in the two-qubit system driven by weak
fields.}  \label{fig:graph2} \end{figure}

As before, first assume that the pulse is described
by a coherent field state, and commute the
displacement operator to the left.
This replaces $\mathcal{E}^\pm$ with
$\mathcal{E}^\pm$ + $E e^{\mp i\bar{\omega}t}$
in Eq. (\ref{eqn:ham1}).
Second, split $H$ = $H_C$ + $H_Q$,
where $H_Q$ contains the electric (or magnetic)
field operators terms.   Suppose we find
$U_C(t)$, the propagator corresponding to $H_C$.
Then the total propagator for the system can be
formally integrated as
\begin{equation}
U(t) = U_C(t) - \frac{i}{\hbar} \int_0^t \, dt' \,
U_C(t-t') H_Q U_C(t') \, -
\frac{1}{\hbar^2} 
\int_0^t dt' \int_0^{t'} dt'' \,
U_C(t-t') H_Q U_C(t'-t'') H_Q U_C(t'') + \cdots
\label{eqn:series} \end{equation}
Computing the value of the first integral
in this series is the goal of this section.
As before, this integral can be physically
interpreted as follows: the qubit evolves
under $U_C$, coherently evolving under the
influence of a classical field until time $t'$,
when a scattering event occurs due to the
quantum nature of the field.   After the scattering,
the qubit / field system evolves as before until
final time $t$.   Higher order corrections then
correspond to multiple scatterings (or, multiple field
operators).   To find $U_C$, 
first assume a solution of the form
$\exp(-i\sum \omega
{\mbox {\Large $a$}}_{\vec{k}} {}^\dag
{\mbox {\Large $a$}}_{\vec{k}} )$
$\exp(-i \bar{\omega} 
[ \mathcal{S}_z + \mathcal{I}_z ]t)$
$U_{C1}$, and then drop the rapidly
oscillating terms.   Then diagonalize
the resulting Hamiltonian,
\[
H_{C1} = \left( \begin{array}{cccc}
( \omega_1 \! + \! \omega_2 )/2 -\bar{\omega}
\! + \! J/4 &
\kappa E/2 & \kappa E/2 & 0 \\
\kappa E/2 &
(\omega_1 \! - \! \omega_2)/2 \! - \! J/4 & 
J/2 & \kappa E/2 \\
\kappa E/2 & J/2 &
(- \omega_1 \! + \! \omega_2)/2 \! - \! J/4 &
\kappa E/2 \\
0 & \kappa E/2 & \kappa E/2 &
-(\omega_1 \! + \! \omega_2)/2 \! + \! \bar{\omega}
\! + \! J/4 \end{array} \right),
\]
to find eigenvalues $\lambda_n$ and 
eigenvectors $v_n$.   Plugging in the $U_C$ 
operator into Eq. (\ref{eqn:series}),
and as before keeping only the terms linear
in the creation operators,
the lowest order quantum back
reaction is (in the rotating frame)
\[
-\frac{\kappa}{2} \,
\sum_{\vec{k}}
e^{i\vec{k}\cdot \vec{r}}
\sqrt{\frac{\hbar \omega}{2\epsilon_0 L^3}} \,
{\mbox {\Large $a$}}_{\vec{k}} {}^\dag
\sum_{n,m=1}^4
v_n v_n^\dag ( \mathcal{S}_- + \mathcal{I}_- )
v_m^\dag v_m
e^{i \lambda_n T}
\frac{\displaystyle e^{i(\omega-\bar{\omega}
+\lambda_m - \lambda_n)t} - 1}
{\displaystyle i(\omega-\bar{\omega}
+\lambda_m - \lambda_n)}
\]
The real parts of the field amplitudes of the
one-photon states entangled with different qubit
states for a few of the matrix elements of
$U$ are shown in Fig. \ref{fig:graph2}.
The parameters used to generate the figure
were $\omega_1$ = 20, $\omega_2$ = 21, 
and $J$ = 0.4 for the qubits, and 
$\kappa E$ = 0.01, and $T$ = 100.6 for
the classical pulse profile.
The classical result, $U_C(T)$, 
transfers 97\% of the
amplitude between states $|11\rangle$
and $|10\rangle$.
These parameters are not meant to
represent any specific spectroscopic method.
\par
For on-resonant transitions, 
the matrix element peaks strongly at
$T$ $\propto$ $E^{-1}$.   Thus, as before,
weaker pulses lead to greater decoherence.
The matrix element for the
transition $|01\rangle \rightarrow |00\rangle$
is found to have the greatest magnitude.
It may be surprising that transitions that
are not being driven by the external field can
also have large decoherence rates.
Recall that this method of implementing
the CNOT gate actually drives all the
allowed transitions, but is only resonant with
one of them.   This decoherence process 
continues to grow because these other transitions
that can lead to photon emission
are weakly but continuously excited.
\par
In summary, several conclusions are of interest.
First, the field states that become entangled
with the two-qubit states are orthogonal
to the initial field state, but they are not mutually
orthogonal.   In fact, their projections onto one
another have essentially random phases.
Computations that use the phases of the qubit
state will be scrambled by this process.
Secondly, because the CNOT gate involves entangling
separate qubits, the scattering process also
involves creating mutually entangled qubit / field
states.   It appears reasonable to suggest that
this kind of field / multiple qubit entanglement
will result when such logic gates are driven by
external fields.
\section{Discussion}
\subsection{Grover's algorithm}
When unitary transforms are driven
by externally generated coherent fields
in the manner discussed above, a decoherence
mechanism exists that, with each applied pulse,
tends to scramble the computer's memory.
This decoherence mechanism is slightly different from
the usual environmentally induced decoherence,
in that it increases as the number of times
the programmer attempts to manipulate the
qubit system coherently.  In the case of
Grover's search algorithm where the Walsh-Hadamard
transforms are externally driven,
the degradation of the correct response 
scales as 
$0.2 K 2^{K/2} 
(\hbar \bar{\omega}/\epsilon_0 L^3) 
(\kappa/\Omega)^2$
\par
A criticism of this analysis might be in
the specific choice used to implement $\mathcal{W}(n)$.
Whatever method is chosen,
the field / qubit propagators still hold,
and some back reaction must exist (but see below).
In general, the degradation should scale 
as the number of times a qubit transform is driven.
For Grover's algorithm,
if no error correction routines are implemented,
then the amplitude $\Omega$ of the field will
have to increase exponentially with increasing $K$ 
in order to keep the error below a fixed bound.
Clearly, this is not a scalable way to implement
Grover's algorithm.
\par
How important is this decoherence mechanism to
the different proposed quantum computer schemes?
Let us employ simple order of magnitude arguments,
and ignore for the moment the implementation of
error-correcting codes.  The total probability
for decoherence can also be written as
0.2
$[ (\hbar \bar{\omega})/(L^3 \epsilon_0 E^2/2) ]$
$K 2^{K/2}$.
The pre-factor is the ratio of the energy
of a single photon, to the total energy in
the pulse (energy density times resonator volume),
or the reciprocal 
of the total number of photons in the
resonator used to create a pulse.
\par
First, examine the case for using lasers to
drive single ions or atoms.   A recent 
experimental demonstration of a logic gate
using trapped $^9\mathrm{Be}^+$
ions as qubits \cite{monroe} 
used 1 mW pulses of $\approx 10^{-4}$ s 
duration at 300 nm.
This corresponds to $10^{12}$ photons per
pulse.   The very small pre-factor will
not pose a problem for computations involving
a polynomial number of steps with increasing $K$,
but for Grover's algorithm this mechanism limits
the number of qubits to $\approx$ 70.
\par
Next, suppose we examine the case for
NMR quantum computing.   First, let us address
the question of utilizing the signal from
a large number of independent quantum computers.
Let us assume that our sample can be prepared
in the ground state (all spins initially
down), and let us ignore the interaction of the
computers with one another through the field.
We note that these assumptions are hardly 
justified for real world systems, but assuming
a finite temperature for bulk quantum computing
causes separate difficulties that have been
addressed elsewhere \cite{warren}.
The application of Grover's algorithm results
in a final state
\[
|\Psi\rangle =
\prod_{\displaystyle i=1}^{\displaystyle 10^{23}}
\bigg( (1-\gamma^2/2)
|y^{(i)}\rangle + \gamma
\sum_x G(x) |x^{(i)}\rangle \bigg)
|\mathrm{vac}\rangle
\]
where $G$ creates orthogonal field states
that contribute little amplitude 
to the correct solution.
The total signal is the sum from all of the
quantum computers in the sample,
$\langle \Psi | 
\sum_i \bigg( |y^{(i)}\rangle
\langle y^{(i)} | \bigg)
| \Psi \rangle$
= N $(1 - \gamma^2/2)$.
The point is that at low temperatures,
the macroscopic decoherence rate is multiplied
by the total number of independent quantum
computers in the sample.
Typically, NMR uses $\nu \approx 10^8$ Hz, 
or a photon energy of $10^{-25}$ J.   
Pulses are 100 W for 10 $\mu$s, 
for a total energy of $10^{-3}$ J.
Thus, there are $10^{22}$ photons per pulse.
This limits Grover's algorithm for NMR to
$\approx$ 140 qubits if we can do NMR on
a single spin system; but for a micro-mole 
($10^{17}$) of quantum computers, 
we are limited to roughly 25 qubits.
If electron spin resonance is substituted as the 
spectroscopic technique, then $\nu \approx 10^{10}$ Hz,
pulses are 1 kW for 10 ns, and thus require $10^{18}$
photons, a negligible improvement over NMR.
\par
If, as is usually the case,
$\Omega \ll \omega_n$, then
the number of photons required to
generate $\mathcal{W}(n)$ 
is proportional to $\Omega / (\kappa^2 \omega_n)$.
Thus, physical systems with 
small values of $\omega_n$ or $\kappa$ are
the most resistant to the above decoherence mechanism.
Unfortunately, such systems have other
limitations: if $\omega_n$ is small,
then the temperature of the system is required
to scale with increasing $K$ in an unfortunate
manner \cite{warren},
while $\kappa$ is small, then the
time required to drive the qubits increases,
which in turn slows the entire computation down.
It seems that quantum computers 
driven or programmed by externally applied
fields will face significant design trade-offs.
\par
Will driving qubits by externally applied, 
static electric fields \cite{kane} 
offer any significant advantages?
(Similar proposals might be envisioned for
NMR by applying different static magnetic fields
along different directions to individual spins.) 
In the Coulomb gauge, such longitudinal fields
are not independent degrees of freedom \cite{cohen}.
Instead, the total Hamiltonian of such a system
must include the particles whose charges give rise
to the static fields.
Thus, for the case of qubits driven by electrostatic
fields from electrodes, decoherence results from
the field operators (now with half integer
spin statistics) for the Fermi levels
of each electrode becoming entangled 
with the qubit operators.
Finding the decoherence rate in this case
is complicated by the extra structure for
the electrons in the electrode, but it would be
surprising if some decoherence mechanism was
not present.
\par
\subsection{Error Correcting Codes}
Finally, we briefly examine some possible
means to reduce the decoherence, and some
difficulties they may encounter because of the
unusual nature of this decoherence mechanism.
First, consider the use of error-correcting codes.
They are a general method by which to 
reduce decoherence \cite{shor,knill},
provided that the decoherence is a 
{\it statistically independent} process 
amongst the qubits within each code-word.
For logic gates applied bit-wise to code-words 
\cite{shor},
these error correcting codes should be able
to correct for this decoherence as for any other
environmentally induced decoherence.
The difficulty comes in when one considers how
to actually implement the detection and correction
scheme.   Because of the requirement that
errors within code-words be independent between
qubits, we can not expect to have qubit / qubit
couplings within each code-word.   For the encoding
schemes for which the authors are aware,
this seems to require the ``ancillary'' bits \cite{shor}
to be driven by external fields in such a way
so that multiple qubits interact with the ancillary
bit at once.   Doing this with external fields,
however, leads to entanglements between the qubits
within the code-word and the external fields.
This in turn means that errors developed during the
error correction scheme itself will not have the
required statistical independence.
However, in order to better understand how 
this decoherence mechanism might influence error
correcting schemes, it would be helpful to examine a
complete program, such as Grover's algorithm
with a specific code implemented.
\par
Perhaps more hopefully, it appears that there
are independent means by which storing information
in multiple qubits could reduce this decoherence
mechanism.   Note that the appearance of terms 
such as $|B_\beta + B_-|$
in Eq. (\ref{eqn:firstdecohere})
shows that destructive interference can lessen
the probability of a back reaction.
It is known to be possible to quench
spontaneous emission in multi-level systems
\cite{scully2}, which is a process quite
similar to the decoherence mechanism considered
here.   Thus, it seems likely that a similar
design could be used to remove the lowest order
terms in the expansion of Eq. (\ref{eqn:series}).
Then, the decoherence rate increases as
the inverse square of the number of photons
in a pulse, which can be considered as 
a significant improvement.
\par
{\large {\bf Conclusions}}
\par
Quantum computers that use external,
classically generated electromagnetic fields
to drive the evolution of the system undergo a
decoherence induced by the quantum back reaction to
those fields.  The probability for the quantum system
to be degraded increases as the total number of
externally driven transforms, and inversely 
as the number of photons in the pulse.
For algorithms in which no error-correcting codes
are implemented, and for which the number of pulses
is required to increase exponentially 
as the problem size increases,
the field intensity driving the system will also
be required to increase exponentially in order to
bound the degradation of the response.
And for the implementation of error-correcting codes,
the use of external fields for error detection and
correction gives rise to a decoherence
that does not have the property of 
being statistically
independent between 
separate qubits in the code-word.
\par
{\large {\bf Acknowledgements}}
\par
We gratefully acknowledge support from
the Air Force Office of Scientific Research.
%
%
\pagebreak \clearpage
%
\pagebreak \clearpage
\end{document}